\documentclass[]{IEEEtran}
\usepackage{cite}
\usepackage{amsmath,amssymb,amsfonts}
\usepackage{algorithmic}
\usepackage{graphicx}
\usepackage{textcomp}
\usepackage{xcolor}

\ifCLASSOPTIONcompsoc
    \usepackage[caption=false, font=normalsize, labelfont=sf, textfont=sf]{subfig}
\else
\usepackage[caption=false, font=footnotesize]{subfig}
\fi

\usepackage{url}								
\usepackage{siunitx}							
\usepackage{booktabs}							
\usepackage{tabularx}
\newcolumntype{Y}{>{\centering\arraybackslash}X}

\usepackage{makecell}							
\usepackage{multirow}							
\usepackage[display-foreign=false]{acro}			
\usepackage{gensymb}
\acsetup{single}								

\usepackage{balance}							

\usepackage{soul}
\usepackage[linewidth=1pt]{mdframed}
\usepackage{layouts}
\usepackage{lipsum}

\newcommand{\hlb}[1]{#1}

\DeclareAcronym{lfm}{
	short = LFM,
	long = linear frequency modulation
}
\DeclareAcronym{prf}{
	short = PRF,
	long = pulse repetition frequency
}
\DeclareAcronym{fmcw}{
	short = FMCW,
	long = frequency modulated continuous-wave
}
\DeclareAcronym{cw}{
	short = CW,
	long = continuous-wave
}
\DeclareAcronym{dbf}{
	short = DBF,
	long = digital beamforming
}
\DeclareAcronym{crlb}{
	short = CRLB,
	long = Cramér-Rao lower bound
}
\DeclareAcronym{dof}{
	short = DoF,
	long = degree of freedom
}
\DeclareAcronym{if}{
	short = IF,
	long = intermediate frequency
}
\DeclareAcronym{adc}{
	short = ADC,
	long = analog to digital converter
}
\DeclareAcronym{fft}{
	short = FFT,
	long = fast Fourier transform,
}
\DeclareAcronym{rmse}{
	short = RMSE,
	long = root-mean-square error
}
\DeclareAcronym{psd}{
	short = PSD,
	long = power spectral density
}

\DeclareAcronym{ni}{
	short = NI,
	long = National Instruments
}

\DeclareAcronym{daq}{
	short = DAQ,
	long = data acquisition
}

\DeclareAcronym{snr}{
	short = SNR,
	long = signal-to-noise ratio
}

\DeclareAcronym{doa}{
	short = DOA,
	long = direction of arrival
}

\DeclareAcronym{adi}{
	short = ADI,
	long = Analog Devices
}

\DeclareAcronym{vco}{
	short = VCO,
	long = voltage-controlled oscillator
}

\DeclareAcronym{lna}{
	short = LNA,
	long = low-noise amplifier
}

\DeclareAcronym{lo}{
	short = LO,
	long = local oscillator
}

\DeclareAcronym{hci}{
	short = HCI,
	long = human-computer interaction
}

\DeclareAcronym{mimo}{
	short = MIMO,
	long = multiple-input multiple-output
}

\DeclareAcronym{simo}{
	short = SIMO,
	long = single-input multiple-output
}

\DeclareAcronym{mmic}{
	short = MMIC,
	long = monolithic microwave integrated circuit
}

\newcommand{\uv}[1] {\hat{\mathbf{#1}}}

\newcommand{\fc}{41.8}

\newcommand{\ExItanVelRmse}{41.01}
\newcommand{\ExItanAngRmse}{10.42}
\newcommand{\ExItanVelMaxDev}{100.27}
\newcommand{\ExItanAngMaxDev}{19.00}

\newcommand{\ExIIradVelRmse}{31.97}
\newcommand{\ExIItanVelRmse}{59.06}
\newcommand{\ExIItrueVelRmse}{45.07}
\newcommand{\ExIIangRmse}{5.11}
\newcommand{\ExIIradVelMaxDev}{66.84}
\newcommand{\ExIItanVelMaxDev}{117.94}
\newcommand{\ExIItrueVelMaxDev}{99.11}
\newcommand{\ExIIangMaxDev}{11.30}

\def\BibTeX{{\rm B\kern-.05em{\sc i\kern-.025em b}\kern-.08em
    T\kern-.1667em\lower.7ex\hbox{E}\kern-.125emX}}
\begin{document}

\title{Three Dimensional Velocity Measurement Using a Dual Axis Millimeter-Wave Interferometric Radar}

\author{ Jason Merlo,~\IEEEmembership{Graduate~Student~Member,~IEEE,} Eric Klinefelter,~\IEEEmembership{Graduate~Student~Member,~IEEE,}\\and Jeffrey A. Nanzer,~\IEEEmembership{Senior Member,~IEEE}%
		\thanks{This material is based in part upon work supported by the Air Force Research Laboratory (contract number FA8650-14-D-1725).}
	\thanks{The authors are with the Department of Electrical and Computer Engineering, Michigan State University, East Lansing, MI 48824 USA (email: merlojas@msu.edu, klinefe4@msu.edu, nanzer@msu.edu).}
}

\maketitle

\begin{abstract}
In this work, a method for directly measuring target velocity in three dimensions using a dual axis correlation interferometric radar is presented.  Recent advances have shown that the \hlb{measurement of a target's angular velocity} is possible by correlating the signals measured at spatially diverse aperture locations.  By utilizing multiple orthogonal baselines and using conventional Doppler velocity methods to obtain radial velocity, a full three-dimensional velocity vector can be obtained using only three receive antennas and a single transmitter, without the need for tracking. A $\text{\fc\,GHz}$ dual axis interferometric radar with a $\mathbf{7.26\lambda}$ antenna baseline is presented along with measurements of a target moving parallel to the plane of the radar array, and of a target moving with components of both radial and tangential velocity. These experiments achieved total velocity \acp{rmse} of $\text{\ExItanVelRmse\,mm}\mathbf{\cdot}\text{s}^{-1}$ ($\mathbf{10.5\%}$) for a target moving along a plane parallel to the array, and $\text{\ExIItrueVelRmse\,mm}\mathbf{\cdot}\text{s}^{-1}$ ($\mathbf{13.5\%}$) for a target moving with components of radial and tangential motion relative to the array; estimated trajectory angle \acp{rmse} of \ang{\ExItanAngRmse} and \ang{\ExIIangRmse} were achieved for each experiment respectively.

\end{abstract}


\acresetall 




\section{Introduction}



\IEEEPARstart{R}{ecent trends} \hlb{in monolithic microwave integrated circuit (MMIC) development have enabled designs of short-range millimeter-wave radars using commodity MMICs instead of discrete components, setting the foundations for ubiquitous radar sensing in everyday devices. } 
Some notable areas of active research in compact radar systems are \ac{hci}, robotics and automotive, and non-contact vital signs monitoring systems.  \Ac{hci} systems primarily employ one of two techniques to capture input: micro-Doppler classification \cite{chen2006micro, chen2019micro}, and multidimensional path reconstruction. Micro-Doppler techniques utilize the micro-motion signatures induced by the differing relative velocities of each part of the hand to classify a certain gesture---these techniques often trade spatial resolution for temporal resolution enabling highly compact systems \cite{molchanov2015multi, lien2016soli, kim2016hand, choi2019short, liang2020gesture}; full multidimensional path reconstruction facilitates positional tracking of fine movements, such as writing in air, but requires higher angular resolution, and thus, larger array aperture sizes \cite{fan2016gesture, zhang2018hci, arsalan2019character}.  In robotics and automotive, recent trends have focused on the improvement of angular resolution to enhance the ability to discriminate between closely separated targets in cluttered environments, as are typically encountered in urban areas, by using a mixture of signal processing-based techniques \cite{fischer2012evaluation, fuchs2019single, zhang2020super}, and physical \cite{gottinger2021coherent, bialer2021super} and synthetic \cite{harrer2017synthetic, feger2017experimental, gishkori2018imaging, tagliaferri2021navigation, iqbal2021imaging} aperture array-based techniques. Non-contact vital signs monitoring utilizes highly sensitive Doppler phase shift measurements obtained by isolating the frequencies induced by minor radial motions of respiration and heartbeat on the surface of the body \cite{li2013review, will2018radar, michler2019clinically}.

While many of the aforementioned radar systems have achieved high levels of performance, it is notable that %
of the six fundamental state variables (range, elevation angle, azimuth angle, range-rate, elevation angle-rate, and azimuth angle-rate),
current radar systems only support direct measurements of the first four,
and lack the ability to make instantaneous measurements of angle-rate in azimuth and elevation \cite{richards2010principles}. 
Presently, many of these systems depend on various types of phased arrays to determine the signal \ac{doa}. \hlb{This enables rapid scanning of the full search space and, for \acf{dbf} phased arrays, enables reception of scattered signals from any direction.} However, they typically require many elements, greatly increasing the cost and complexity of the system.  
These systems also require the angular rate of the target to be estimated by taking multiple angle estimates of the target over time, tracking the target, then computing its angular velocity, which is not desirable when compared with a direct measurement.  These systems not only require \ac{doa} algorithms such as the commonly used variants of MUSIC \cite{schmidt1986music} or ESPRIT \cite{roy1989esprit}, but they also require tracking and data association algorithms \cite{blackman1986tracking, bogler1990radar}. This creates increased computational requirements, increased uncertainty due to indirect measurements, and cannot obtain angular velocity information on the order of radar fast-time.
However, it has recently been demonstrated that an interferometric technique which correlates the signals received at two spatially disparate  apertures, similar to that used in radio astronomy \cite{richard2017interferometry}, can be used to obtain the angular velocity of a target directly via frequency estimation, analogous to the measurement of radial velocity via Doppler shift estimation \cite{nanzer2010mtt, klinefelter2019chapter}.

In this paper, we demonstrate for the first time, a \ac{simo} radar system capable of directly measuring the velocity of targets in three dimensions (radial, azimuth angular, and elevation angular velocities). Extensively expanding our prior work \cite{merlo2020dualaxis}, we show that the traditional Doppler velocity measurement can be combined with the aforementioned angular velocity measurement and extended to two orthogonal axes to obtain an estimate of the full, instantaneous target velocity vector in three dimensions, without the need for tracking, thereby reducing computational requirements and latency over alternative approaches. Whereas most previous works have only demonstrated the measurement of angular velocity in one dimension, this work is the first to demonstrate simultaneous measurement of the three-dimensional velocity vector based on knowledge of the target's position (range and angle). The target range and angle were previously shown to be directly measurable using this technique by utilizing a modulated transmit waveform {\cite{vakalis2018angle, merlo2020joint}}. \hlb{Alternatively, differential Doppler and phase-based tracking techniques \cite{armstrong1998target, ristic2012doppler_tracking, battistelli2013new, fan2016gesture}} could be also used to obtain the target position, which may  benefit from the direct angular velocity measurement afforded by this technique to improve their positional estimates. \hlb{However, for} the purposes of this paper, the target's trajectory is assumed to be known a priori via one of these or other traditional radar approaches. \hlb{Although the techniques in {\cite{ristic2012doppler_tracking, battistelli2013new, fan2016gesture, armstrong1998target}} are capable of measuring the angular velocity of a target using {\ac{cw}} systems via differential tracking means}, the key difference between these techniques and our proposed method is the correlation of the received signals, enabling direct angular velocity measurement via frequency estimation without requiring tracking of the target. 


The organization of this paper is as follows. The fundamental equations for angular velocity measurement are discussed, followed by an extension of these equations into multiple dimensions, in Section \ref{fundamentals}. Section \ref{sec:resolution-and-accuracy} discusses the resolution and lower bound on accuracy for the methods used in this paper. The \SI{\fc}{\giga\hertz} dual axis interferometric radar hardware is presented in Section \ref{radar_hardware}. Finally, experimental measurements and simulations using the interferometric radar are presented for both fully tangential trajectories and trajectories with components of both radial and tangential velocity in Section \ref{experiments}. We demonstrate the ability to estimate the angular velocity of reflecting targets with \acp{rmse} of $\text{\ExItanVelRmse\,mm}\mathbf{\cdot}\text{s}^{-1}$ and $\text{\ExIItrueVelRmse\,mm}\mathbf{\cdot}\text{s}^{-1}$ for each experiment respectively, with trajectory angle estimation \acp{rmse} of \ang{\ExItanAngRmse} and \ang{\ExIIangRmse} respectively.

\section{Interferometric Radar Measurement of Angular Velocity}
\label{fundamentals}
\begin{figure}[tb]
  \includegraphics{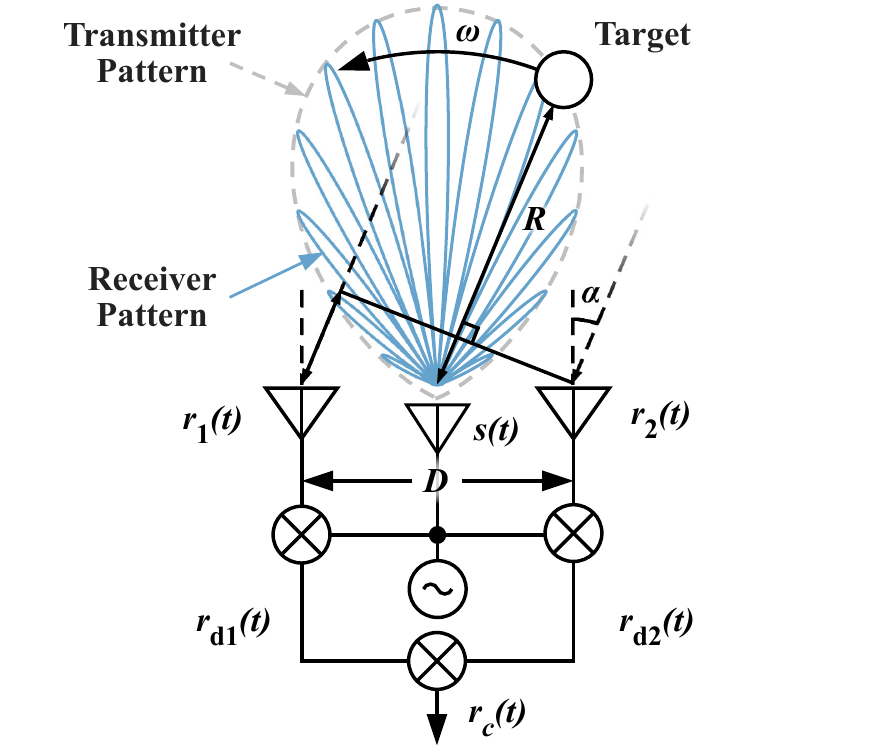}
  \caption{Schematic of a single axis slice of an active direct-downconversion correlation interferometer using semi-directive antennas. The transmitter antenna pattern is indicated by the grey dotted line and the fringe pattern generated by the receive antennas is represented by the solid blue line. Multiple baselines may share the same transmitter.}
  \label{schematic}
\end{figure}

The angular velocity of a target may be directly measured by estimating the frequency produced by the correlation of the signals received at two spatially separated antennas as the target moves past the aperture of the array \cite{nanzer2010mtt, klinefelter2019chapter}. The received signal may be intrinsically emitted by the target (via thermal \cite{nanzer2010mtt} or other means) or it may be a reflected signal from a transmitter \cite{klinefelter2019chapter}. The correlation of the signal received at two receiving antennas with a large electrical spacing is modulated by the time-changing phase of the signal as it passes through the grating lobe pattern created by the widely spaced apertures, as shown in Fig. \ref{schematic}.

For an active \ac{cw} radar system the signal received at each antenna after downconversion, neglecting the initial phase and loss terms, is described by
\begin{equation}
	r_{d,n} \propto e^{-j2\pi f_0\tau_{d,n}}
\end{equation}
where $f_0$ is the center frequency of the tone and $\tau_{d,n}$ is the round-trip time delay of the signal to antenna $n$. After being received and downconverted, the two antenna signals are correlated yielding 
\begin{align}
	\label{r_c}
	\begin{split}
		r_{c} &\propto r_{d,1} \cdot r_{d,2}^* \\
		&\propto e^{-j2\pi f_0(\tau_{d,2}-\tau_{d,1})}\\
		&\propto e^{-j2\pi f_0 \frac{D}{c} \sin{\alpha}}
	\end{split}
\end{align}
where $D$ is the antenna separation distance, $c$ is the speed of light in the medium, and $\alpha$ is the angle of the target off broadside. \hlb{The time derivative of the complex exponential's phase in \eqref{r_c} is} obtained to determine the instantaneous frequency response of the interferometer due to the angular motion of a target. By characterizing the time-dependent angle $\alpha$ by the target angular velocity about the center of the array $\omega$ to \hlb{produce} $\alpha = \omega t$, the instantaneous frequency shift due to the angular velocity of the target \hlb{is found as}
\begin{equation}
	\label{ang_vel_full}
	f_\omega = \omega {D}_{\lambda}\cos{\alpha}
\end{equation}
where $D_\lambda=Df_0/c$ is the baseline distance between the antennas in wavelengths. 

If the target is near broadside, or if the transmit or receive antennas have a sufficiently narrow beam pattern, a narrow-beam approximation is commonly used (i.e., $\cos\alpha\approx1$) further reducing \eqref{ang_vel_full} to $f_\omega=\omega D_\lambda$. Thus, if the distance to the target $R$ is known (which can be estimated via a traditional radar measurement or other means), the tangential linear velocity is found by applying the relation $v = \omega R$ resulting in 
\begin{equation}
	\label{ang_vel_small}
	v_\alpha = \frac{f_\omega R}{D_\lambda}
\end{equation}
which is valid for small angles (the error from this approximation is~$<10\%$~for~$|\alpha|\le\ang{20}$).


\begin{figure}[tb]
  \includegraphics{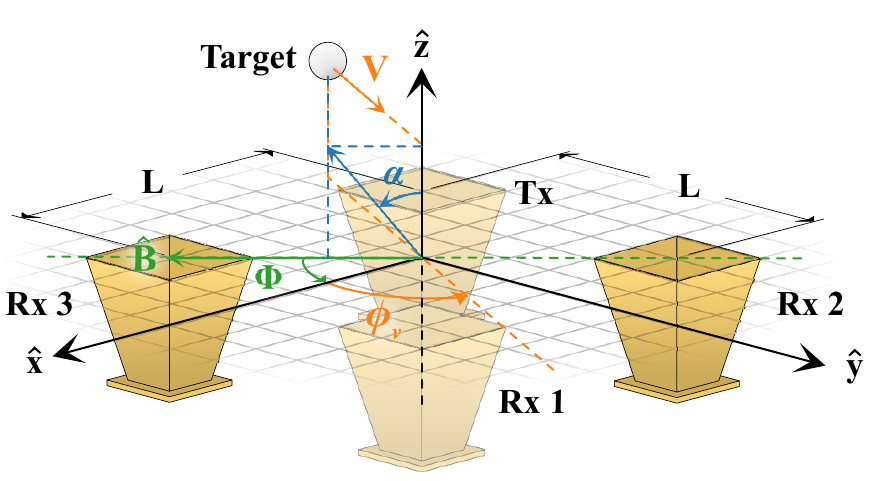}
  \caption{Positioning of the transmit and receive horns, and measurement coordinate system. The vector $\mathbf{V}$ is the velocity vector of the target. The vector $\hat{\mathbf{B}}$ is the vector between the receiving elements in the active baseline. The angle $\alpha$ is the elevation angle of the target projected onto the plane formed by the active baseline vector $\hat{\mathbf{B}}$ and $\hat{\mathbf{z}}$. The angle $\phi_v$ is the angle between the velocity vector $\mathbf{V}$ and $\hat{\mathbf{x}}$. The angle $\Phi$ is between the active baseline vector $\hat{\mathbf{B}}$ and $\hat{\mathbf{y}}$. For baselines $\Phi=\ang{0},\,\ang{90}$, $D=L$, however for $\Phi=\ang{-45}$ as shown in this figure, $D=\sqrt{2} L$.}
  \label{array-coordinates}
\end{figure}

While prior works have focused primarily on single axis measurements, as depicted in Fig. \ref{schematic}, we extend this concept to the measurement of the three-dimensional velocity vector by employing two orthogonal baselines, as illustrated in Fig. \ref{array-coordinates}; this is accomplished by simultaneously measuring angular velocity in two dimensions, while incorporating the traditional Doppler measurement of radial velocity.  In this case, a single \ac{cw} transmitter is shared by two baselines arranged in a square pattern with side length $L$; note that for the orthogonal baselines $D=L$, however for the \ang{-45} baseline between Rx2 and Rx3, the baseline distance is $D=\sqrt{2} L$. As is common with correlation interferometry systems, $D$ should be chosen to be multiple wavelengths (typically $D>5\lambda$) to ensure the phase at the output of the correlator will process through multiple cycles as the target passes the array, improving the ability to estimate the frequency induced by the tangential motion of the target at the output of the correlator. Finally, in the dual axis configuration, $\alpha$ is generalized further as the angle of the target projected onto the plane formed by the $z$-axis of the array and the baseline vector drawn between the antennas on a given baseline $\uv{B}$ as shown in Fig. \ref{array-coordinates}.

Because the interferometer only measures the component of angular velocity about the axis orthogonal to its baseline, for targets moving off-axis, the expected interferometer response, given the narrow-beam approximation, will be proportional to \eqref{ang_vel_small} multiplied by the cosine of the angle between the baseline vector and the component of the target velocity vector projected onto the normal surface of the array,
\begin{equation}
	\label{3d_ang_vel}
	v_\alpha = \frac{f_\omega R}{D_\lambda}\cos{\left(\phi_v + \Phi\right)}
\end{equation}
where $\phi_v$ is the angle between the $x$-axis and the velocity vector projected onto the normal surface of the array, and $\Phi$ is the angle between the baseline vector $\uv{B}$ and the antennas used for the measurement, and the $x$-axis and $\phi_v,\,\Phi\in[\ang{-90},\,\ang{90}]$.

In the three-dimensional case, the velocity vector is represented by $\mathbf{V} = \left<v_R,\, v_\phi,\, v_\theta \right>$ with the components representing the radial and tangential velocities in the azimuthal and elevational directions, respectively.  The first of these quantities can be directly obtained by measuring the Doppler frequency shift $f_d$ at one, or multiple antennas using $v_R = f_d \lambda / 2$, and averaging the results. The second and third components of the velocity vector may be derived using the interferometric response from the two orthogonal baselines. It should be noted that each axis of the interferometer will only measure the component of tangential velocity about the axis orthogonal to its baseline vector, thus for clarity, these will be denoted as $v_{\alpha_x}$ and $v_{\alpha_y}$, referring to the component of angular velocity measured by the $\hat{\mathbf{x}}$ and $\hat{\mathbf{y}}$ baselines, respectively, found using \eqref{3d_ang_vel}.

\begin{table}[tb]
	\caption{Table of Variables}
	\label{tab:variables}
  	\begin{tabularx}{\columnwidth}{p{0.1\linewidth} p{0.8\linewidth}}
	
	\toprule[1pt]
	\multicolumn{2}{l}{\textbf{Radar Measurement Quantities}} \\
	\midrule
	\midrule
	
	Symbol & Definition \\
	\midrule
	$R$ & Range from array to target \\
	$\theta$ & Elevation angle between zenith and target \\
	$\phi$ & Azimuth angle relative to $x$-axis \\
	$v_{\alpha_x}$ & Component of angular velocity measured on the $\Phi=\ang{0}$ baseline \\
	$v_{\alpha_y}$ & Component of angular velocity measured on the $\Phi=\ang{90}$ baseline \\
	$v_R$ & Component of velocity towards the center of the array \\
	
	\midrule[1pt]
	\multicolumn{2}{l}{\textbf{Derived Target State Quantities}} \\
	\midrule
	\midrule
	
	Symbol & Definition \\
	\midrule
	$\phi_v$ & Target track azimuth angle with respect to target center \\
	$v_\theta$ & Tangential velocity of the target with respect to the array \\
	$\beta$ & Target angle of attack relative to the plane of the surface of the array \\
	
	\bottomrule[1pt]
	\end{tabularx}
\end{table}

The components of velocity measured by each axis can be described by
\begin{align}
	\label{full-3d-equations}
	\begin{split}
		v_{\alpha_x} &= v_\phi \cos{\phi} + v_\theta \cos{\phi} \cos{\theta} \\
		v_{\alpha_y} &= v_\phi \sin{\phi} + v_\theta \sin{\phi} \cos{\theta}
	\end{split}
\end{align}
where $\phi$ and $\theta$ are the azimuth and elevation of the target, respectively, and $\phi=0$ is in the $\uv{x}$ direction, while $\theta=0$ is the zenith.
The magnitude of the tangential components can then be found by taking the $L^2$ norm of the azimuthal and elevational components of the velocity. \hlb{From \mbox{\eqref{full-3d-equations}} it is shown} that to obtain the full three-dimensional velocity of the target for an arbitrary position in space, the target's elevation and azimuth must be estimated.  In prior work, it has been shown that using a modulated transmit waveform, the same principles of correlating the signals received at multiple spatially distributed receivers can be used to determine the target's range and angle \cite{vakalis2018angle, merlo2020joint}; however, this work focuses on the direct three-dimensional velocity measurement of targets, thus to simplify the localization problem, the target's trajectory is taken to be known a priori.  In these experiments, the target is placed on a linear guide which passes directly through the zenith of the array with a known constant velocity enabling the determination of the target's range $R$ and elevation $\theta$ at any given time. Thus, to determine the target's full three-dimensional trajectory, only three additional quantities must be estimated: the heading azimuth angle of the target passing through the zenith of the array $\phi_v$,  the magnitude of the target's tangential velocity $v_\theta$, and the target angle of attack relative to the surface of the array $\beta$. A summary of the measured and derived quantities and their descriptions is provided in \mbox{Table \ref{tab:variables}}. 



Once the components of tangential velocity about each axis are measured, the angle of the target trajectory can be found using
 \begin{equation}
	\label{phi_v}
	\phi_v = \tan^{-1}{\left(v_{\alpha_y}/v_{\alpha_x}\right)}.
\end{equation}
Next, the magnitude of tangential velocity may be found using the $L^2$ norm of the component axes
\begin{equation}
 	\label{v_theta}
	v_\theta = \sqrt{v_{\alpha_y}^2+v_{\alpha_x}^2}.
\end{equation}
Finally, the angle of attack, or elevation angle of the trajectory $\beta$ can be found by evaluating the inverse tangent of the radial velocity and the magnitude of the tangential motion from \eqref{v_theta} resulting in
 \begin{equation}
	\label{beta}
	\beta = \tan^{-1}{\left(-v_R/v_\theta\right)}.
\end{equation}
In this case, $\beta\in[\ang{-90},\,\ang{90}]$, where $\beta=\ang{0}$ represents a trajectory parallel to the plane of the array.

Measurement of the three-dimensional velocity vector using the dual axis interferometric radar thus depends on three frequency estimations: the Doppler frequency shift, which may be obtained from a Fourier analysis of one or more of the radar receivers, and the azimuthal and elevation interferometric frequency shifts, which may be obtained by Fourier analysis of the \hlb{correlated signals from the orthogonal baseline pairs.}

While the implications of adding tangential measurements to the set of \hlb{fundamental radar measurements} should not be understated, it is apparent from \eqref{r_c} that when multiple targets are present, the received signals will be the sum of all scatterers present. Thus, the correlation step will perform a mixing operation between all pairs of signals producing $N^2$ responses for $N$ dynamic targets, which is not desired. While mitigation of this intermodulation between the targets is not the focus of this work, it is important to highlight some of the work and techniques proposed to perform this mitigation.  In \cite{nanzer2014distortion} the use of a very long wavelength carrier was proposed to make the effect of the Doppler shift negligibly small, effectively causing the intermodulation terms to collapse to the same frequency as the desired correlation tones after the time averaging correlation process. Alternatively, \cite{nanzer2014distortion} also proposed the use of very short pulses to eliminate the temporal overlap of scattered signals in the correlator from targets at different ranges resulting in elimination of distortion for targets with non-overlapping range bins. In \cite{wang2019simultaneous}, the use of an inverse Radon transform was proposed as a specialized method for dealing with the intermodulation by reconstructing the signal using the desired target's motion signature, in that case, rotating bodies. More recently, the authors of \cite{wang2020transversal} proposed a more general method utilizing a uniform linear array of three receivers to perform spatial interferometric averaging of the outputs of the adjacent correlation pairs to cancel the frequencies introduced by the radial motion, leaving only the frequency corresponding to the angular motion of the targets.  Finally, in \cite{merlo2021multiple} we proposed and demonstrated another general method utilizing an interferometer consisting of multiple diverse baseline lengths. This exploited the fact that the desired responses on each baseline will align when scaled by each respective baseline's length; intermodulation terms will be scaled arbitrarily, thus allowing mitigation by simply multiplying the spectra from all normalized baselines.

\section{Resolution and Accuracy of the Interferometric Measurement}
\label{sec:resolution-and-accuracy}

The ability of a radar system to resolve targets closely separated in velocity is an important practical consideration.
The Doppler resolution of a radially moving target has been studied extensively and is given by the half-power bandwidth of the spectral response of a time-limited \ac{cw} signal\cite{richards2010principles}
\begin{equation}
	\label{beam-pattern}
	\delta f_{D}=\frac{1}{T},
\end{equation}
where $T$ is the observation time. The resolution of a correlation interferometer has been explored in depth for the measurement of angle rate for both active and passive systems in \cite{nanzer2012resolution}. The result is still fundamentally a frequency resolution problem, similar to the Doppler resolution; this is apparent from the form of \eqref{ang_vel_full}: given the narrow-beam assumption, the frequency at the output of the interferometer is proportional to the angular velocity, thus the resolution is again dependent on the half-power bandwidth of the response. For an active \ac{cw} system, the signal bandwidth of the interferometer can be obtained by finding the Fourier transform of the system response to a moving point target \eqref{r_c} windowed by its antenna pattern; for a practical system, the main beam of the antenna pattern can be approximated by a Gaussian function 
\hlb{
\begin{equation}
	\label{beam-pattern}
	A(\theta)=\exp\left(-\frac{\theta^2}{2\sigma^2}\right)
\end{equation}}%
whose half-power beamwidth is approximated by \mbox{$\theta_{BW}\approx2.355\sigma$}. Using this, the frequency response of the system is found to be
\begin{equation}
	\label{beam-pattern}
	R_c(f)=\sqrt{\frac{2\pi\sigma^2}{\omega^2}}\exp\left[-\frac{2\pi^2\sigma^2}{\omega^2}\left(f_0-f_\omega\right)^2\right]
\end{equation}
resulting in a minimum interferometric resolution of the system given by
\begin{equation}
	\label{resolution}
	\delta f_{int}=2.355\frac{\omega}{\pi\theta_{BW}}
\end{equation}
meaning that resolution may be improved by reducing the angular rate of the target $\omega$ or by increasing the system beamwidth $\theta_{BW}$; both results are intuitively verifiable as both will cause the effective observation time to increase, resulting in a narrower spectral response.

The accuracy of a measurement is a similarly important practical parameter of a radar system. While the effective accuracy of a system is dependent on the system itself and the estimator implemented, a lower-bound on accuracy may be obtained in the form of the \acf{crlb} which is informative in determining the performance of a system relative to its maximum achievable theoretical performance. \hlb{The lower bound on accuracy of the Doppler shift estimate of a radially moving target is known to be \cite{nanzer2017accuracy, richards2010principles}
\begin{equation}
	\label{doppler-accuracy}
	\text{var}\left(\tilde{f}_d-f_d\right)\ge\frac{N_0}{2|\eta|^2\left(\zeta_t^2-\frac{1}{E_s}\mu^2_t\right)}
\end{equation}
where $f_d$ is the true Doppler frequency shift and $\tilde{f}_d$ is its estimated value; $N_0$ is the noise power; $\eta$ is the \hlb{complex amplitude due to the channel and scattering}; $\zeta_t^2$ is the mean-square mean time duration of the signal; $\mu_t$ is the mean time duration of the signal; $2\left|\eta\right|^2/N_0$ is the peak \ac{snr};} and $E_s \triangleq \int \left | s \left( x; \tau,\,f_d \right)\right|^2dx$ where $\tau$ is the time delay of the received signal. This corresponds to a radial velocity measurement accuracy of
\begin{equation}
	\label{radial-velocity-accuracy}
	\text{var}\left(\tilde{v}_R-v_R\right)=\frac{c^2}{4f_0^2}\text{var}\left(\tilde{f}_d-f_d\right).
\end{equation}
The accuracy of the angular velocity component has been studied in depth in \cite{nanzer2017accuracy, sharp2017accuracy} and is shown to take a very similar form to the Doppler measurement
\hlb{
\begin{equation}
	\label{angular-frequency-accuracy}
	\text{var}\left(\tilde{f}_\omega-f_\omega\right)\ge\frac{N_0}{2|\eta|^2\left(\zeta_A^2-\frac{1}{E_s}\mu^2_A\right)}
\end{equation}
where $\zeta_A^2$ is the mean-squared antenna pattern of the system, or the second moment of the antenna pattern; $\mu_A$ is the mean antenna pattern of the system, or the first moment of the antenna pattern, which is nominally zero for two-element interferometers due to symmetry; and $E_s \triangleq \int \left | A(\theta)\right|^2d\theta$ where $A(\theta)$ is the antenna pattern.} Finally, assuming $\alpha$ is small, this can be translated to a bound on tangential velocity using
\begin{equation}
	\label{angular-velocity-accuracy}
	\text{var}\left(\tilde{v}_\alpha-v_\alpha\right)=\frac{R^2}{D_\lambda^2}\text{var}\left(\tilde{f}_\omega-f_\omega\right).
\end{equation}

\section{Dual Axis Millimeter-Wave Interferometric Radar System}
\label{radar_hardware}

\begin{figure}
	\includegraphics{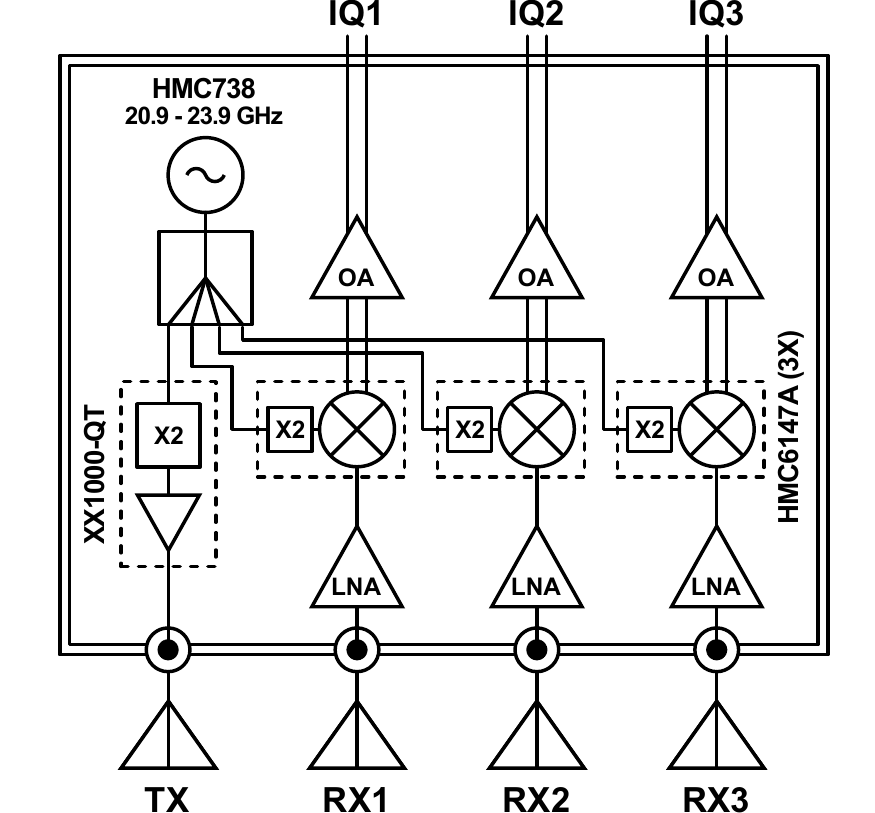}
	\caption{System schematic for the three channel, direct-downconversion, \SI{\fc}{\giga\hertz} \ac{cw} radar used in this experiment. Antennas were connected offboard via \SI{2.4}{\milli\meter} connectors.}
	\label{radar_schematic}
\end{figure}
\begin{figure}
	\includegraphics[width=\columnwidth]{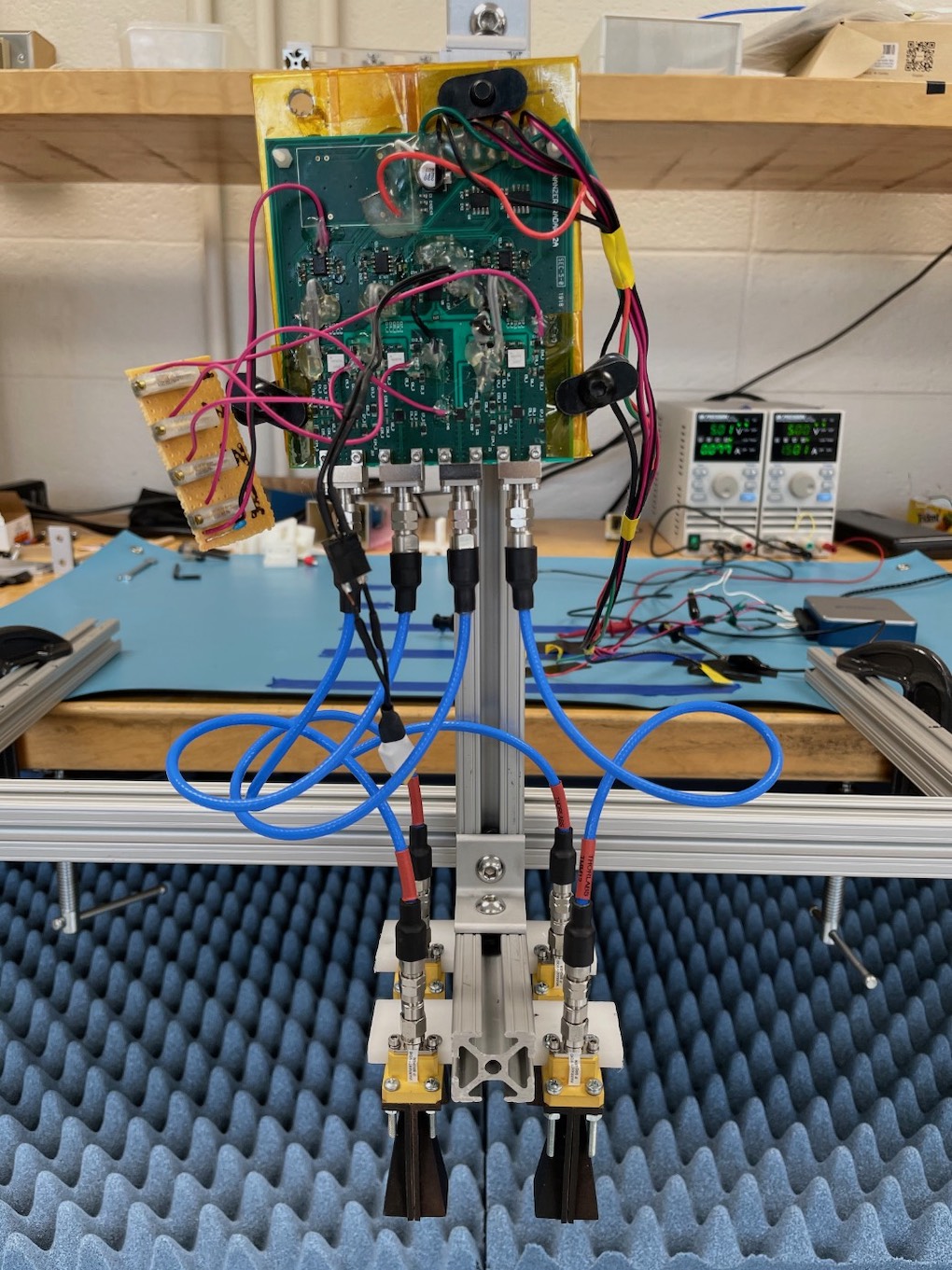}
	\caption{Photograph of the four channel radar printed circuit board connected to a 2x2 array of horn antennas. The $15\,$dBi 3-D printed horns were used for the three receive channels and the transmit channel.}
	\label{radar_detail}
\end{figure}

The radar system used in this experiment utilized one \ac{cw} transmitter operating at \SI{\fc}{\giga\hertz} and three direct-downconverting receivers.  A schematic of this radar system is shown in Fig. \ref{radar_schematic}.  The system was designed with three receive channels to allow two orthogonal receive antenna baselines to be implemented, enabling multidimensional angular velocity measurements. The antennas were connected to the board via \SI{2.4}{\milli\meter} connectors and low-loss cables to allow for the antennas to be placed in different configurations; in this experiment, the antennas were arranged in a square pattern with side length $L=7.26\lambda$.  The transmit and receive antennas were $15\,$dBi 3-D printed metalized horn antennas, fabricated at Michigan State University.

The signal chain on the board consisted of a $20.9$--$23.9\,$GHz \ac{adi} HMC738 MMIC \acf{vco} followed by a 4:1 Wilkinson power divider which split the signal to the transmitter and three downconverters. The transmit path contained a single MACOM \mbox{XX1000-QT} GaAs pHEMT active doubler which directly drove the transmit antenna; the measured transmit power at the antenna was +2.66\,dBm. Each of the three receive paths contained an \ac{adi} HMC1040 GaAs pHEMT \acf{lna}, with a noise figure of $2.2\,$dB and a gain of $23\,$dB, followed by an \ac{adi} HMC6147A downconverter which contained an integrated frequency doubler on the \acf{lo} port. Finally, the baseband signal at the output of the downconverters was amplified by an operational amplifier in an inverting configuration with a gain of $23\,$dB. The baseband signals were sampled by an NI \mbox{USB-6002} \acs{daq} at \SI{4.167}{\kilo Sps}. Signal processing was implemented offline in Python.

The board was fabricated on Rogers RO4350 \SI{0.508}{\milli\meter} substrate with $\epsilon_r=3.71$.  The 1:4 Wilkinson power splitter was designed and simulated in Ansys HFSS and achieved a total insertion loss of $7\,$dB.  The power splitter and associated circuit network, which operated from \mbox{$20.25\,$-$\,22.50\,$GHz}, were designed using microstrip transmission lines, while the other transmission lines operating from \mbox{$40.5\,$-$\,45.0\,$GHz} were designed as grounded coplanar waveguides.  The final design was simulated using Ansys SIwave to ensure that the transmission line $S_{11}$ values were below $-10\,$dB, as well as to ensure uniform impedance. The fabricated board with antennas \hlb{is shown} in Fig. \ref{radar_detail}.

\section{Experimental Results}
\label{experiments}
\begin{figure}[tb]
  \includegraphics{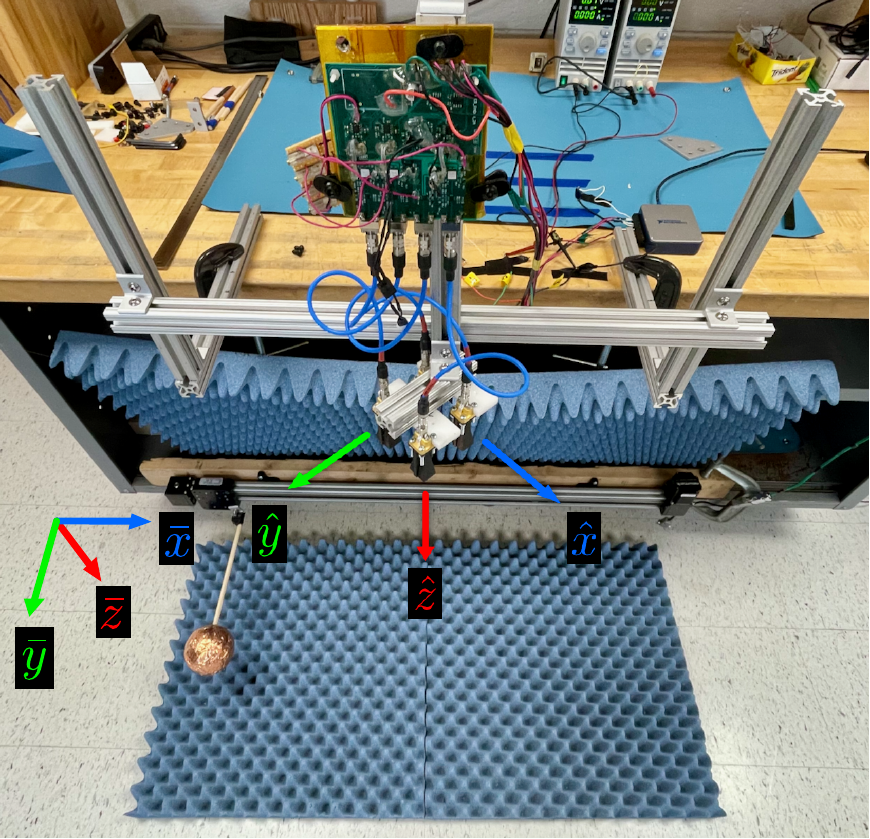}
  \caption{Experimental configuration. A computer controlled linear guide was used to move the metallic sphere target through the field of view of the radar. The array coordinate system is shown around the array in the center and the static coordinate system used for describing target motion is denoted in the lower left corner.}
  \label{experimental-configuration}
\end{figure}

Two experiments were conducted to measure the three-dimensional velocity of a target.  The first experiment measured the angle \hlb{between the target's heading and the} $x$-axis of the array $\tilde{\phi}_v$ as well as the \hlb{speed} of the target while the angle of the array was varied about its $z$-axis, as defined in Fig. \ref{array-coordinates}. The second experiment measured the elevation angle $\tilde{\beta}$ and magnitude of the target velocity $\|\tilde{\mathbf{V}}\|$ moving with velocity components in both the radial and tangential directions for varying elevation angles of the radar.
The target used for the experiments was a \hlb{{\SI{10}{\centi\meter}} diameter polystyrene sphere} coated in copper tape, affixed with a spruce wood dowel to a computer-controlled linear guide with a velocity resolution of \SI{0.54}{\milli\meter\cdot\second^{-1}}. To describe the motion of the target, a fixed coordinate system is used ($\bar{x},\,\bar{y},\,\bar{z}$); nominally, this system aligns with the coordinates used on the array, however the array position is varied throughout the experiments. In each measurement, the guide was set to a maximum speed of \SI{501.31}{\milli\meter\cdot\second^{-1}} and first traversed in the $+\bar{x}$ direction, then back in the $-\bar{x}$ direction. The experimental setup and static coordinate system are shown in Fig. \ref{experimental-configuration}. To observe the impact of variation due to noise, for each measured angle in the experiments, 50 passes of the target were measured in each direction totaling 100 measurements per angle; an analysis of this data was then performed to present the statistical variation across repeated measurements for the chosen processing parameters.  Finally, the observed \ac{snr} across all experiments was between \hlb{16--27\,dB.}

\subsection{Tangential Velocity}
\label{2d-tangential-experiment}

\begin{figure}[tb]
  \includegraphics{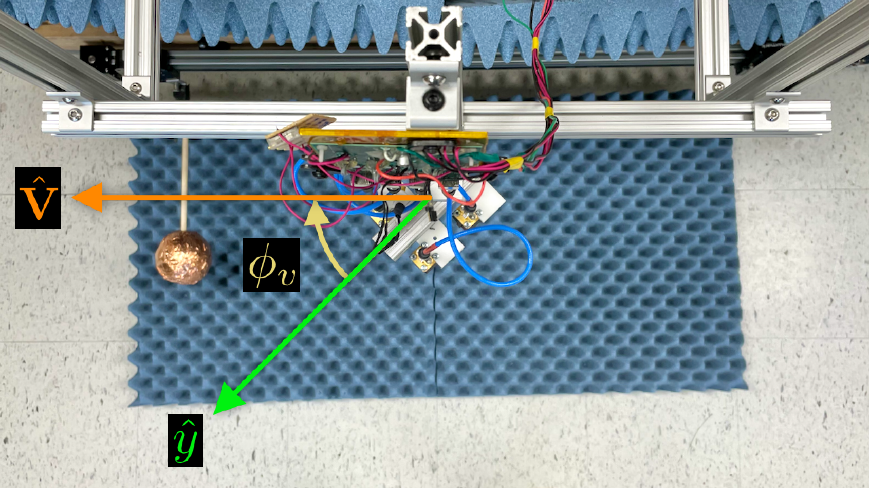}
  \caption{Tangential velocity measurement experimental configuration and the coordinate system.}
  \label{bearing-setup}
\end{figure}

The tangential velocity experiment measured the azimuth angle $\tilde{\phi}_v$ and \hlb{speed} of the spherical target $\|\tilde{\mathbf{V}}\|$ moving parallel to the plane of the radar array. In this case, the Doppler shift is zero at the point of closest approach to the array; due to the narrow-beam pattern of the antennas used, this results in a predominantly tangential motion measurement. For all passes, the distance from the top of the sphere to the approximate center of the array was \SI{755}{\milli\meter}, equal to the far-field distance to the array, at the point of closest approach. To simulate the azimuth component of the target velocity changing in a controlled manner, the antenna array was rotated approximately about its $z$-axis as shown in Fig. \ref{bearing-setup}. Measurements were taken on the $\Phi=\left\{-45, 0,\, 90\right\}^\circ$ baselines with the true azimuth angles of the array set to $\phi_v=\left\{0,\, -15,\, -30,\, -45\right\}^\circ$.

\begin{figure}[tb]
	\centering
	\noindent
	\vspace{-1mm}	
	\subfloat[]{%
	\includegraphics[width=3.4in]{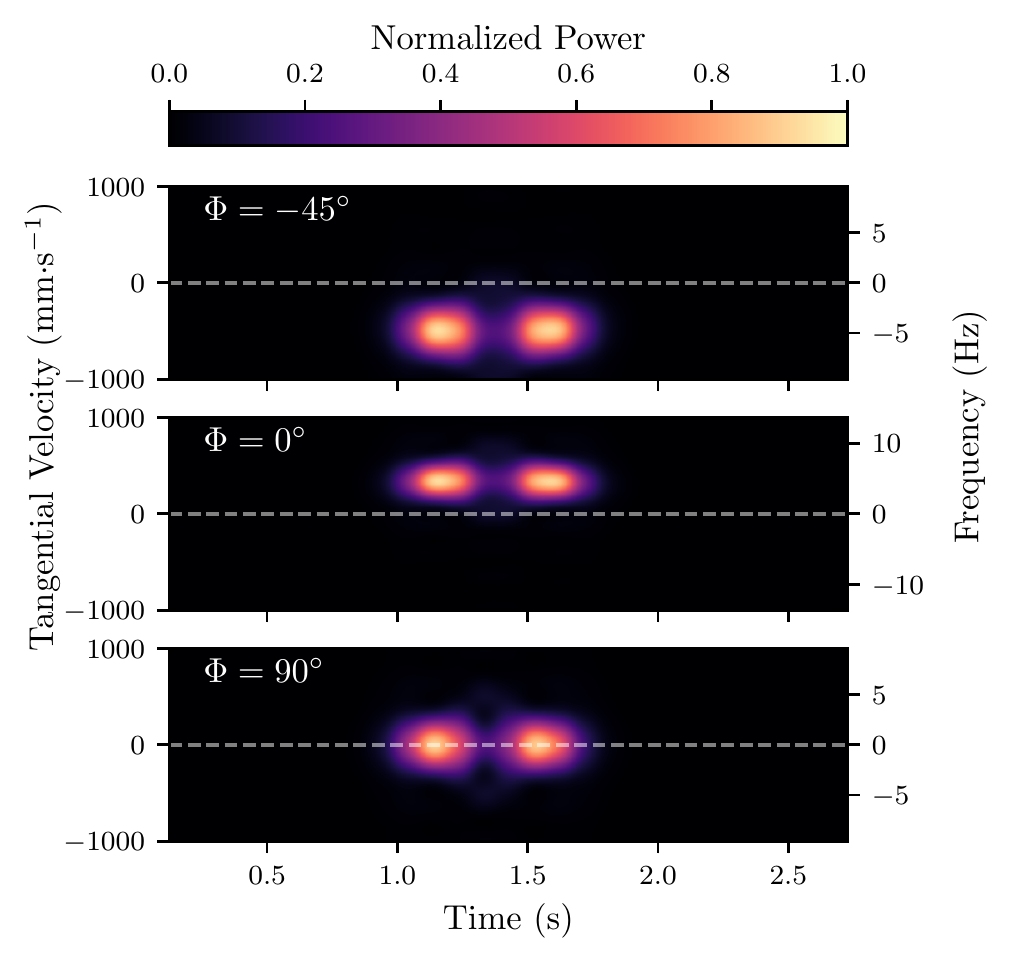}
	\label{sim-bearing-int}}
	\\\vspace{-2mm}
	\subfloat[]{%
	\includegraphics[width=3.4in]{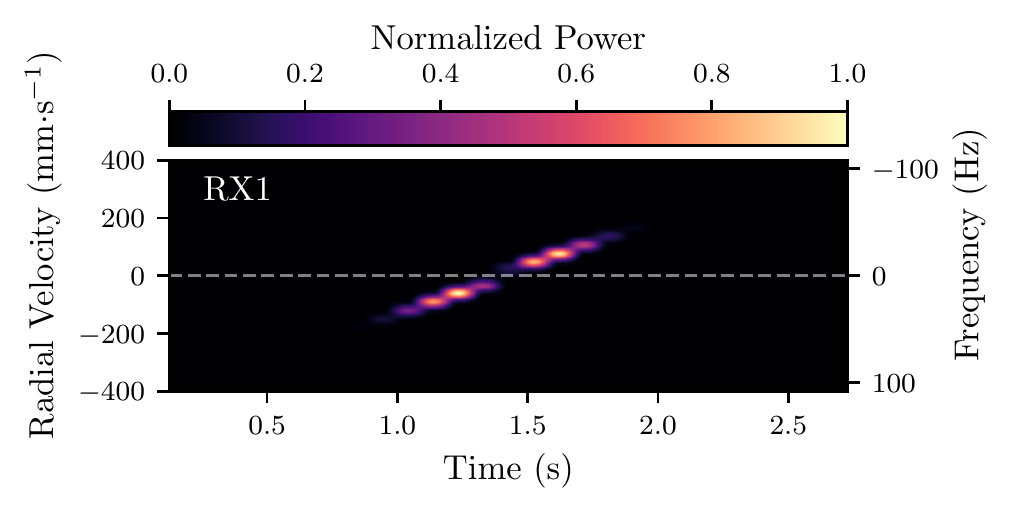}
	\label{sim-bearing-doppler}}
	\caption{(a) Spectrograms of the simulated interferometric response at $\Phi=\ang{0}$ and $\Phi=\ang{90}$, and (b) the simulated Doppler response for the pass of the sphere in the $+\bar{x}$ direction with a speed of \SI{501.31}{\milli\meter\cdot\second^{-1}}, and with $\phi=\ang{0}$ and $\beta=\ang{0}$.}  
	\label{sim-azimuth-spectral}
\end{figure}

\begin{figure}[tb]
	\centering
	\noindent
	\vspace{-1mm}	
	\subfloat[]{%
	\includegraphics[width=3.4in]{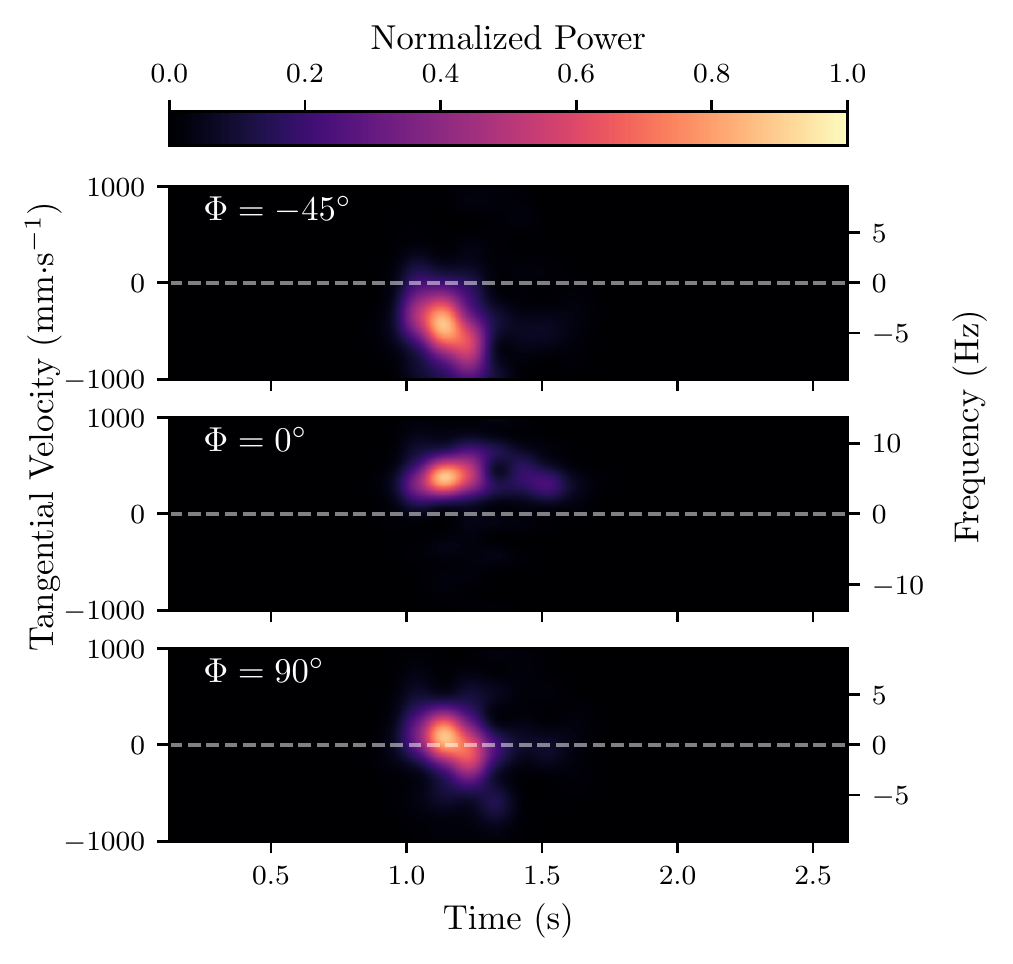}
	\label{bearing-int}}
	\\\vspace{-2mm}
	\subfloat[]{%
	\includegraphics[width=3.4in]{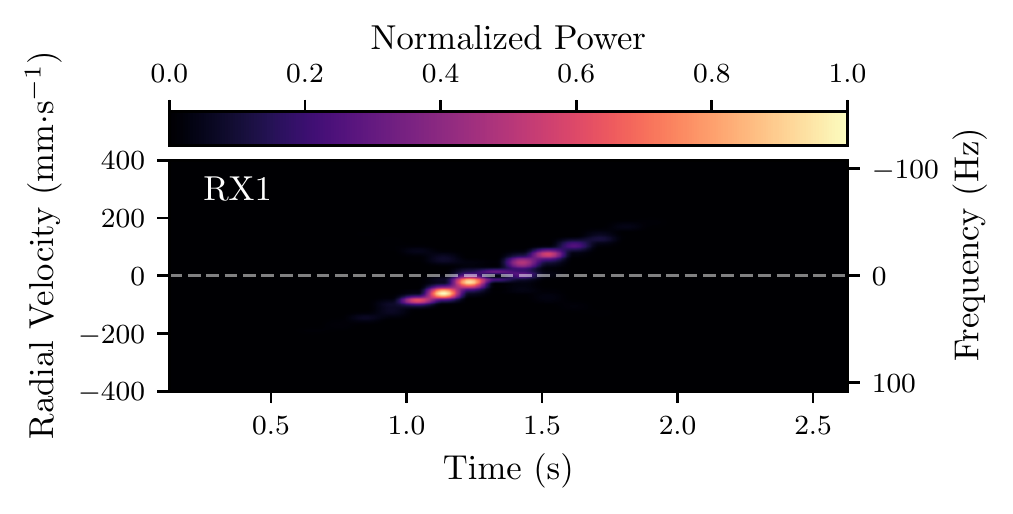}
	\label{bearing-doppler}}
	\caption{(a) Spectrograms of the measured interferometric response at $\Phi=\ang{0}$ and $\Phi=\ang{90}$, and (b) the Doppler response for the pass of the sphere in the $+\bar{x}$ direction with a speed of \SI{501.31}{\milli\meter\cdot\second^{-1}} with $\phi=\ang{0}$ and $\beta=\ang{0}$. The second peak observed when the target passes broadside is significantly weaker in the measurements because the transmitter was offset, increasing the reflected power when the target was approaching; this is not observed in the simulation where the transmitter was ideal.}
	\label{azimuth-spectral}
\end{figure}

\begin{figure*}[tb]
  \centering
  \includegraphics{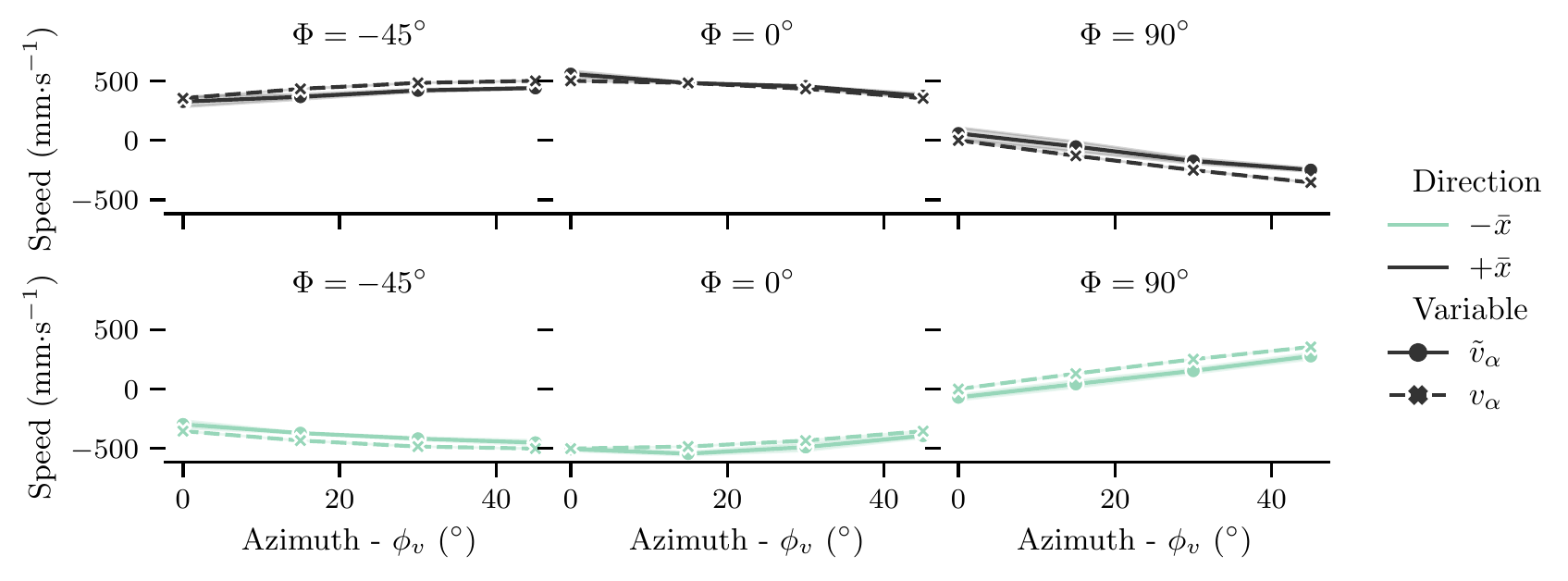}
  \caption{Measured target speed on each baseline across the range of $\phi_v$. The dashed lines indicate the modeled values, and the solid lines indicate the measured values. The shaded region around the lines in the figures indicates the standard deviation of the 50 samples in each direction.}
  \label{correlation-speed}
\end{figure*}

\begin{figure*}
\centering
\includegraphics{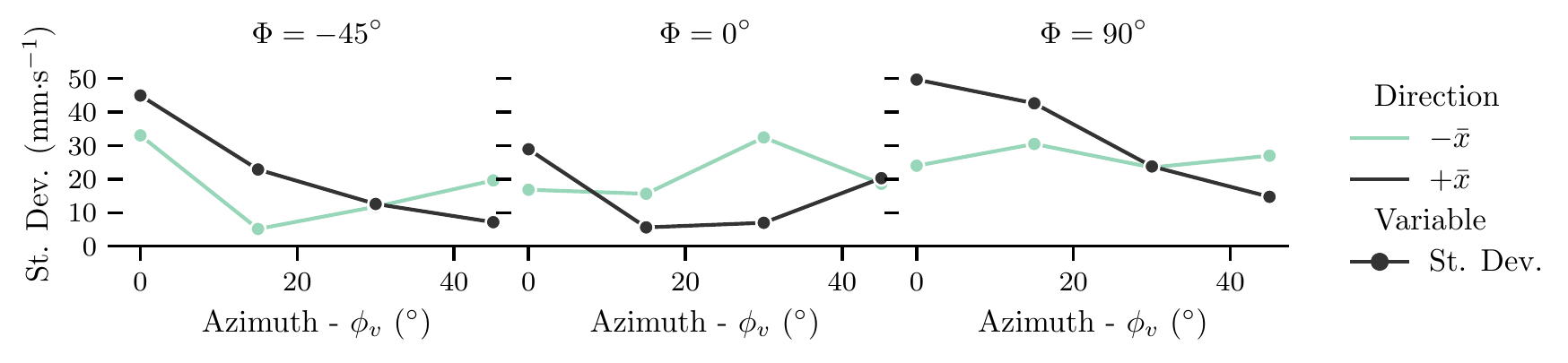}
\caption{\hlb{Measured target tangential velocity standard deviation for each of the baselines at each azimuth angle $\phi_v$.}}
\label{correlation-crlb}
\end{figure*}

To validate the expected interferometric and Doppler spectral responses of an ideal point target passing the array in the nominal orientation, a simulation using a model of the dual axis interferometer \hlb{used in these experiments} was performed. \hlb{The target parameters, array dimensions,} and antenna beamwidths were identical to the experimental configuration.  The simulated correlation and Doppler responses \hlb{are shown} in the spectrograms in Figs. \ref{sim-bearing-int} and \ref{sim-bearing-doppler}, respectively.
The interferometric and Doppler plots were produced using \acp{fft} of size $2^{14}$ with a boxcar window length of \SI{250}{\milli\second} and a $60\%$ overlap.  While an increased overlap was found to improve frequency estimation, the increased lag produced an unacceptable bias in the measurement.
The velocity of the target was estimated on each pass by first estimating the time of closest approach based on the peak \ac{psd} of the Doppler channel 1, then using that time to find the frequency corresponding to the peak \acp{psd} of each of the correlation baselines and of the Doppler shift  on channel 1. 
It should be noted that the $\Phi=\ang{-45}$ baseline has a baseline distance of $D=\sqrt{2}\cdot 7.26\lambda$ due to the array geometry, thus the frequency response is higher by a factor of $\sqrt{2}$ (as \hlb{shown} on the right axis of Fig. \ref{sim-bearing-int}) and must be scaled accordingly when computing the angular velocity. The measured interferometric and Doppler spectral responses of the spherical target for the same array configuration are shown in Figs. \ref{bearing-int} and \ref{bearing-doppler}, respectively. The measured results were processed using the same process and parameters as the simulated results.

Based on the spectral plots it \hlb{is shown} that as the azimuth angle between the target \hlb{heading} and the baseline moves from $|\Phi|=\ang{0}$ to $|\Phi|=\ang{90}$ (for $\phi_v=0$), the frequency response diminishes.  This is expected because the interferometer measures the angular velocity of the target about the axis orthogonal to its baseline described by the cosine term in \eqref{3d_ang_vel}, which is the rate of change of $\alpha$ as depicted in Fig. \ref{array-coordinates}, i.e., the angle projected onto the plane \hlb{formed by} the active baseline and $z$-axis. Thus, when the baseline is parallel to the direction of motion, as in the $\Phi=\ang{0}$ case, the estimated velocity is nearest to the true velocity of the target; when the baseline angle is at an offset of $\Phi=\ang{-45}$ the estimated target velocity magnitude will be $1/\sqrt{2}$ times the true velocity; when the baseline is completely orthogonal to the direction of motion of the target, as in the $\Phi=\ang{90}$ case, the estimated velocity will, ideally, be zero because the phases received at each antenna as the target passes below the array will be identical. In practice, however, using a direct downconversion radar architecture produces a null when \hlb{the} target transitions from moving towards to away from the radar as it passes the array, which is clearly visible in the Doppler response in Figs. \ref{sim-bearing-doppler} and \ref{bearing-doppler}. Furthermore, low frequencies are often attenuated by system high-pass characteristics making low Doppler shifts difficult to detect. Thus, when performing peak \ac{psd} estimation, a time slightly off the time of closest approach is obtained.  To correct for this time shift in the experiment, the known target velocity and direction were used to determine the shift in angle and radius \hlb{off of ideal} which were used to calibrate for this offset. This effect \hlb{is visible} in the spectrograms in Figs. \ref{sim-azimuth-spectral} and \ref{azimuth-spectral}, as the peaks appear to be bimodal as opposed to a unimodal peak as would generally be expected.  It should also be noted that in the measured example in Fig. \ref{azimuth-spectral}, the bimodal peak is significantly less visible than in the simulated shown in Fig. \ref{sim-azimuth-spectral}. This is due to the position of the transmit antenna relative to the receiving antennas; the scattered power received was significantly larger on the side of the array containing the transmitter which causes this peak to appear much more intense.

\begin{figure}[tb]
  \includegraphics[width=\columnwidth]{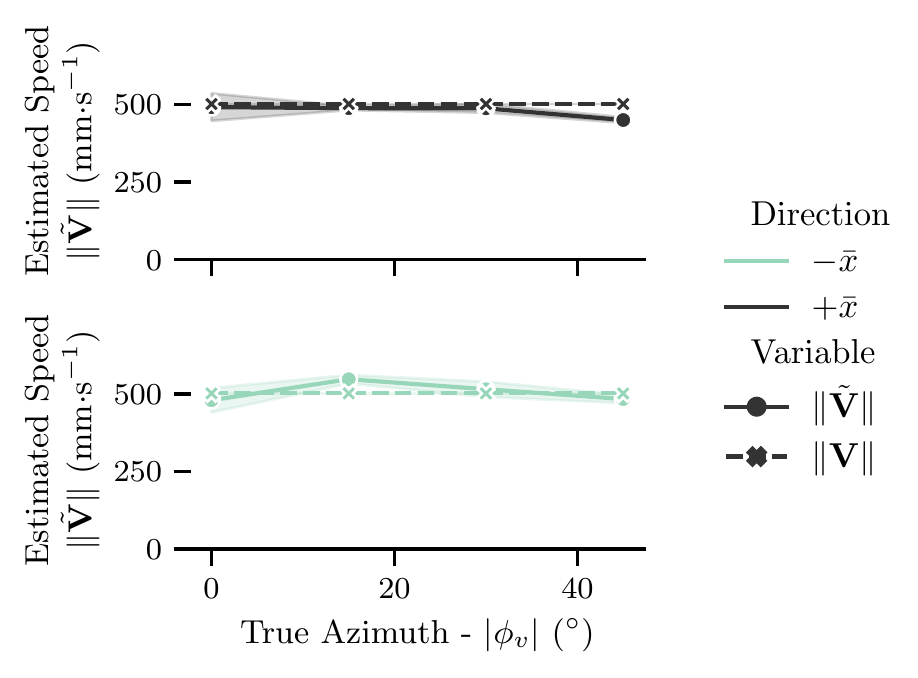}
  \caption{Estimated \hlb{speed} while varying the true target azimuth heading angle ($\phi_v$) in the tangential velocity experiment. The solid lines represent the estimated values while the dashed line represent the true values. The shaded region around the lines in the figures indicates the standard deviation of the 50 samples in each direction.}
  \label{bearing-velocity}
\end{figure}

\begin{figure}[tb]
  \includegraphics[width=\columnwidth]{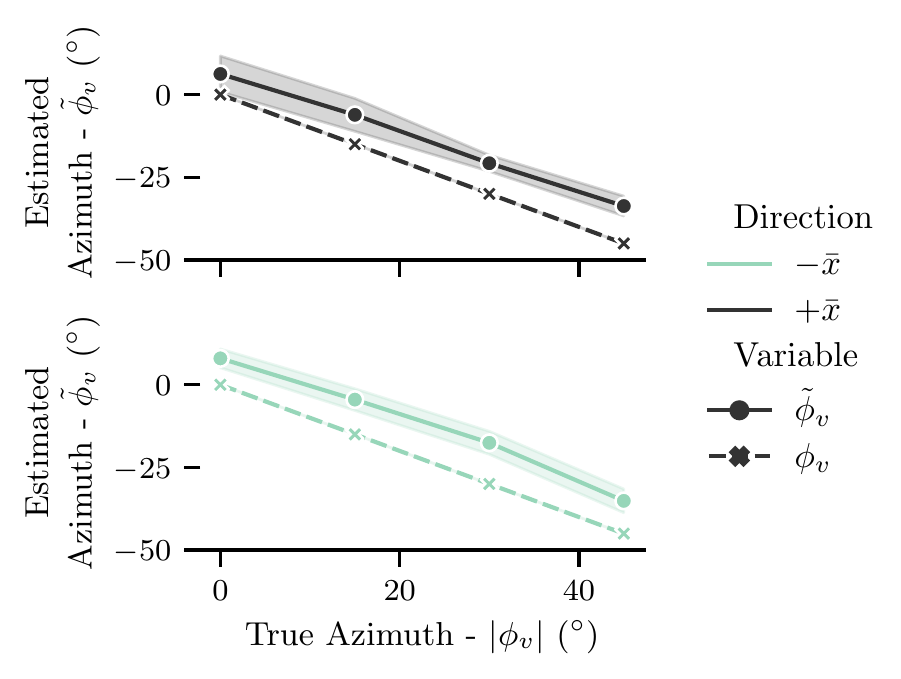}
  \caption{Estimated vs. true azimuth heading angles ($\phi_v$) in the tangential velocity experiment. The solid lines represent the estimated values while the dashed lines represent the true values. The shaded region around the lines in the figures indicates the standard deviation of the 50 samples in each direction.}
  \label{bearing-angle}
\end{figure}

\begin{table}[tb]
	\caption{Tangential Velocity Experiment Estimate Errors}
	\label{tab:ExI_error}
  	\begin{tabularx}{\linewidth}{*{5}{Y}}
	
	\toprule
 
	\multicolumn{2}{c}{Quantity} & Symbol & RMSE & Max. Error \\

	\midrule
	
	\multicolumn{2}{c}{Velocity (\SI{}{\milli\meter\cdot\second^{-1}})} & $\tilde{v}_\theta$ & \ExItanVelRmse & \ExItanVelMaxDev \\
	\multicolumn{2}{c}{Angle ($^\circ$)} & $\tilde{\phi}_v$ & \ExItanAngRmse & \ExItanAngMaxDev \\

	\bottomrule
	\end{tabularx}
\end{table}

The speed measured on each axis is of the interferometer across the range of $\phi_v$ is shown in Fig. \ref{correlation-speed}.  The trend following the cosine curves as described by \eqref{3d_ang_vel} is visible throughout the sweep on each baseline. These component axis values were then used to compute \hlb{both the the speed, using the $L^2$ norm of the $\Phi=\ang{0}$ and $\Phi=\ang{90}$ baselines, and the azimuth heading angle, using \eqref{phi_v}.} The results of this experiment are summarized in Figs. \ref{bearing-velocity} and \ref{bearing-angle}, showing the estimated (solid lines) and expected (dashed lines) target speeds and azimuth heading angles, respectively. The shaded regions around the lines in the figures represent the standard deviation across the 50 measurements taken in each direction for each angle of $\phi_v$. The \acp{rmse} and maximum deviations for the peak spectral estimates \hlb{are shown} in Table \ref{tab:ExI_error}.
The tangential velocity magnitudes for the measurements in Fig. \ref{bearing-velocity} were obtained using the interferometric velocities measured by the \mbox{$\Phi=\ang{0}$} and \mbox{$\Phi=\ang{90}$} baselines and computed using \eqref{v_theta}. Across the range of true velocity azimuth angles $\phi_v$, the estimated target \hlb{speed} had an \ac{rmse} of \SI{\ExItanVelRmse}{\milli\meter\cdot\second^{-1}}.
The estimated target \hlb{heading} angles shown in Fig. \ref{bearing-angle} were estimated using the same measured interferometric velocities from baselines $\Phi={0,\, 90}^\circ$ and computed using \eqref{phi_v}. Over the range of azimuth angles, the estimated values had an \ac{rmse} of $\ang{\ExItanAngRmse}$. 

While the estimated speed shows very good agreement with the true value of the linear guide, the azimuth angle estimate displays a consistent bias of $\sim$\ang{10}.  This is believed to be due to a minor rotational misalignment of the support material used to mount the transmit antenna and receiver 3 causing the baselines formed with receiver 3 to shrink slightly and rotate; this is apparent in both the $\Phi=\ang{-45}$ and $\Phi=90$ baselines in Fig. \ref{correlation-speed} as both show biases consistent with a rotation of their respective axes. As a result, the quantities measured on these axes are no longer orthogonal which introduces error when computing the inverse tangent in \eqref{phi_v}. 

\hlb{Finally, to illustrate the performance of the system during this experiment, the angular velocity accuracy on each baseline of the interferometer at each azimuth angle is presented in Fig. \ref{correlation-crlb}. The mean standard deviation across all correlation measurements for this experiment was \SI{22.43}{\milli\meter\cdot\second^{-1}}. The theoretical mean \ac{crlb} across all azimuth angles, computed using \eqref{angular-velocity-accuracy}, was \SI{0.31}{\milli\meter\cdot\second^{-1}}. The measured standard deviation is greater due in part to the necessity of first estimating the target position off of broadside due to the null as the target passes directly below the array, as previously discussed. This additional uncertainty is not captured by \eqref{angular-velocity-accuracy}. Thus, the \ac{crlb} indicates that the angular velocity measurement can be further improved over the presented results with a more refined position estimator; however, for comparison, the measured standard deviation is of similar magnitude to the $\sim$\SI{10}{\milli\meter\cdot\second^{-1}} achieved by \cite{armstrong1998target}, though it should be noted that the technique used in \cite{armstrong1998target} is designed only for ballistic trajectories.}

\subsection{Tangential and Radial Velocity}
\label{3d-tangential-radial-experiment}

\begin{figure}[tb]
  \includegraphics{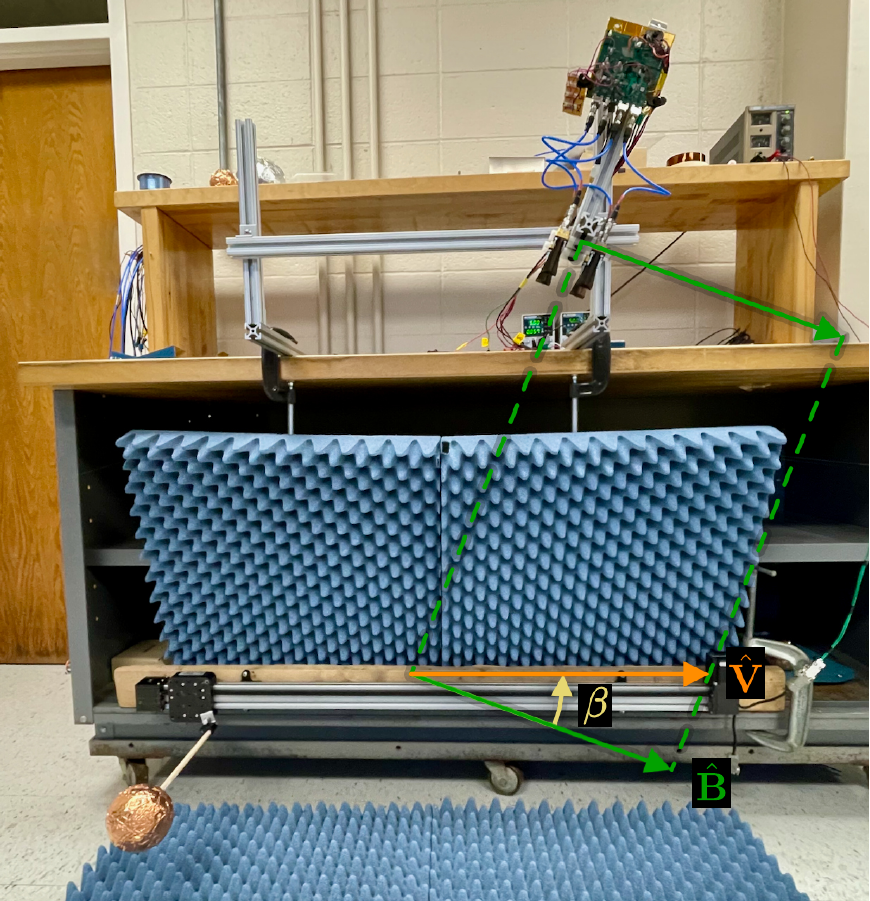}
  \caption{Elevation measurement experimental configuration and coordinate system.}
  \label{elevation-setup}
\end{figure}

The tangential and radial velocity measurement experiment was designed to measure the velocity components of a moving target with a non-zero component of radial motion relative to the array.  To achieve this, the radar array's horizontal location and inclination angle $\beta$ were varied along a slotted aluminum rail as shown in Fig. \ref{elevation-setup}. The elevation angle was varied over $\beta=\left\{0,\,10,\,20,\,30,\,40\right\}^\circ$ while the angular and radial velocities were measured by the array. For each measurement, the angular velocity was estimated using the measurements from the $\Phi=\{0,\,90\}^\circ$ baselines, and the radial velocity was measured at receiver 1.

\hlb{To} measure the target once it was at its full velocity, and not accelerating, the radar array was translated horizontally along the slotted aluminum track as the angle was adjusted to ensure that the $z$-axis of the array was aligned with the center of the linear guide as is shown in Fig. \ref{elevation-setup}.  As a result of this translation, the radius of the measurements $R$ at the time of measurement was also varied throughout the experiment from $R=\SI{755}{\milli\meter}$ to $R=\SI{917}{\milli\meter}$, which was \hlb{accounted for} when computing the \hlb{target's} tangential velocity. As previously noted, the range measurement can be obtained using a typical ranging waveform and can be used in conjunction with the correlation process to additionally obtain the angular position of the target. 

\begin{figure}[tb]
  \includegraphics[width=\columnwidth]{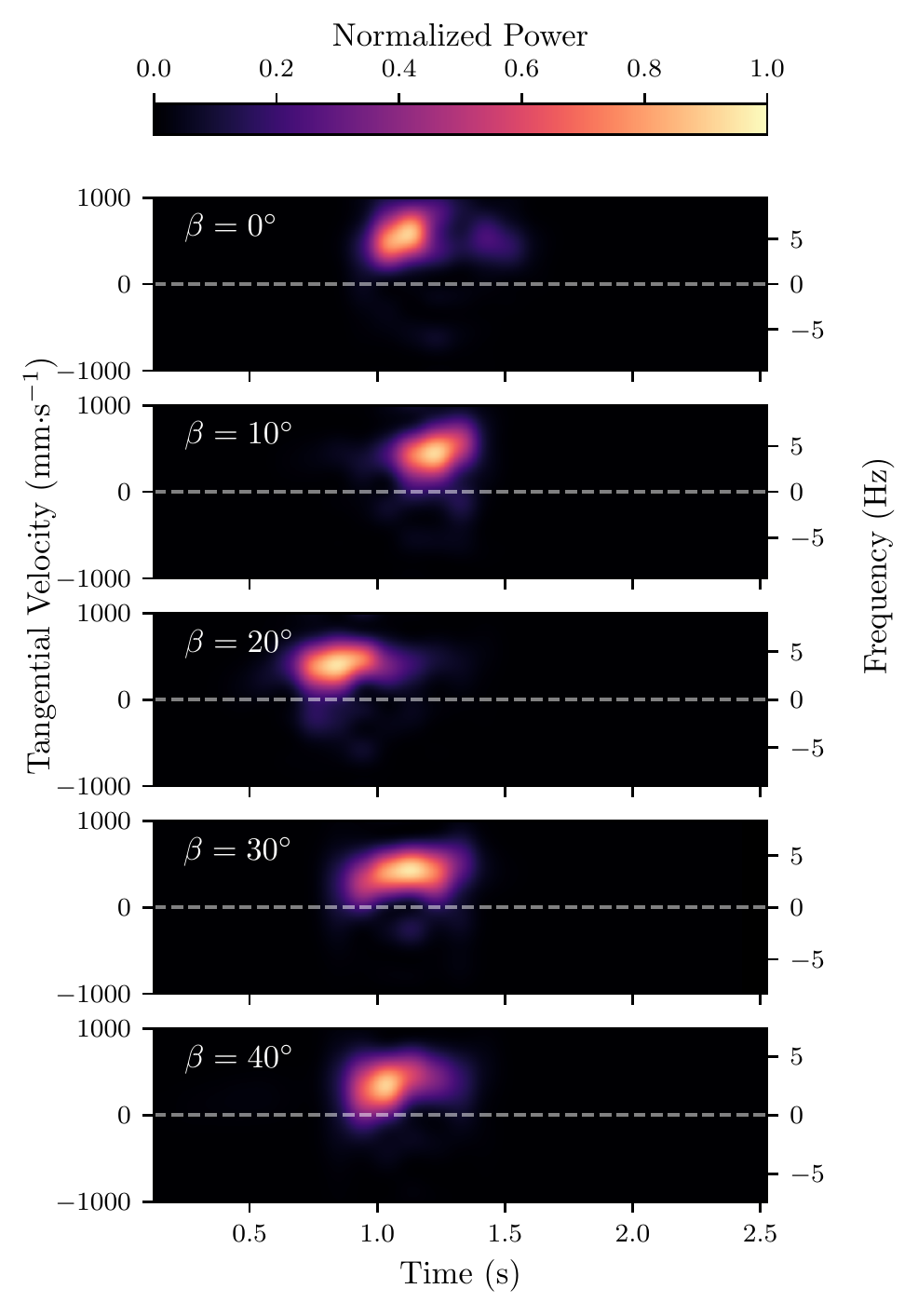}
  \caption{Spectrograms of the measured interferometric responses on the \mbox{$\Phi=\ang{0}$} baseline for varying elevation angles $\beta$ as the sphere passed in the $\tilde{\mathbf{x}}$ direction with a speed of \SI{501.31}{\milli\meter\cdot\second^{-1}}.} 
  \label{elevation-measurement}
\end{figure}

\hlb{The spectral plots of the angular velocity measurements} about the $x$-axis of the array \hlb{are shown} for each angle of $\beta$ in Fig. \ref{elevation-measurement}.  The parameters used to generate the spectrograms and estimate the angular velocity of the target were the same as those described in Section \ref{2d-tangential-experiment}.
When viewing the spectral plots sequentially as $\beta$ increases, two areas should be highlighted. First, as $\beta$ increases, the component of angular velocity begins to decrease, which is expected because target is moving more radially towards the radar, decreasing its tangential component proportionally to the cosine of $\beta$. Second, it should be noted that when the target passes through the $z$-axis of the array, for non-zero values of $\beta$, the correlation process produces unimodal peaks as expected \hlb{because} the target's Doppler shift imparted by the radial velocity no longer crosses the zero Hertz threshold.




\begin{figure}[tb]
  \includegraphics[width=\columnwidth]{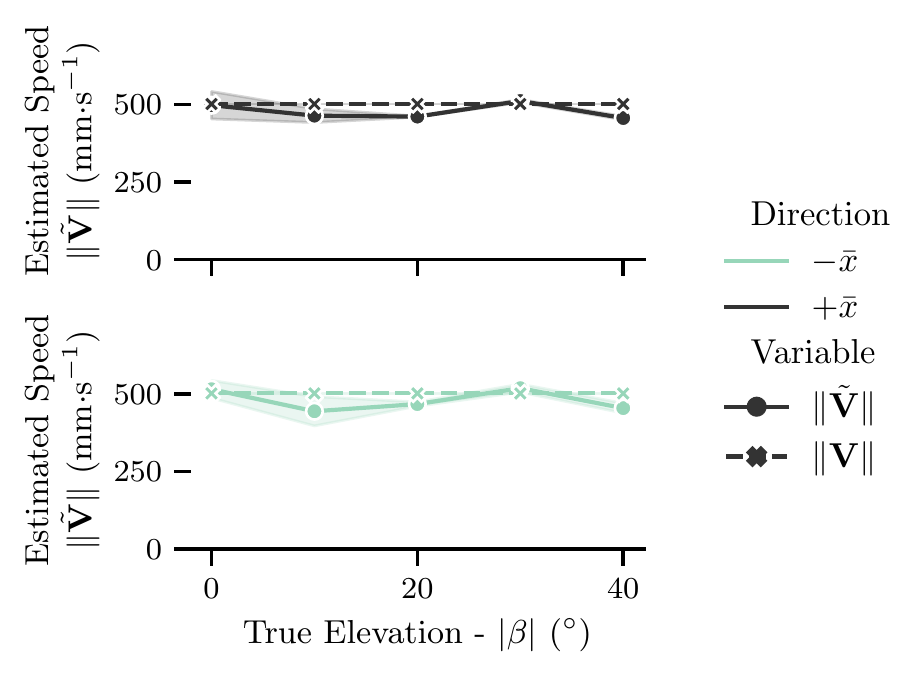}
  \caption{Estimated and true target speed vs. elevation angle in the tangential and radial velocity experiment. The solid lines represent the estimated values while the dashed lines represent the true values. The shaded region around the lines in the figures indicates the standard deviation of the 50 samples in each direction.}
  \label{elevation-velocity}
\end{figure}

\begin{figure}[tb]
  \includegraphics[width=\columnwidth]{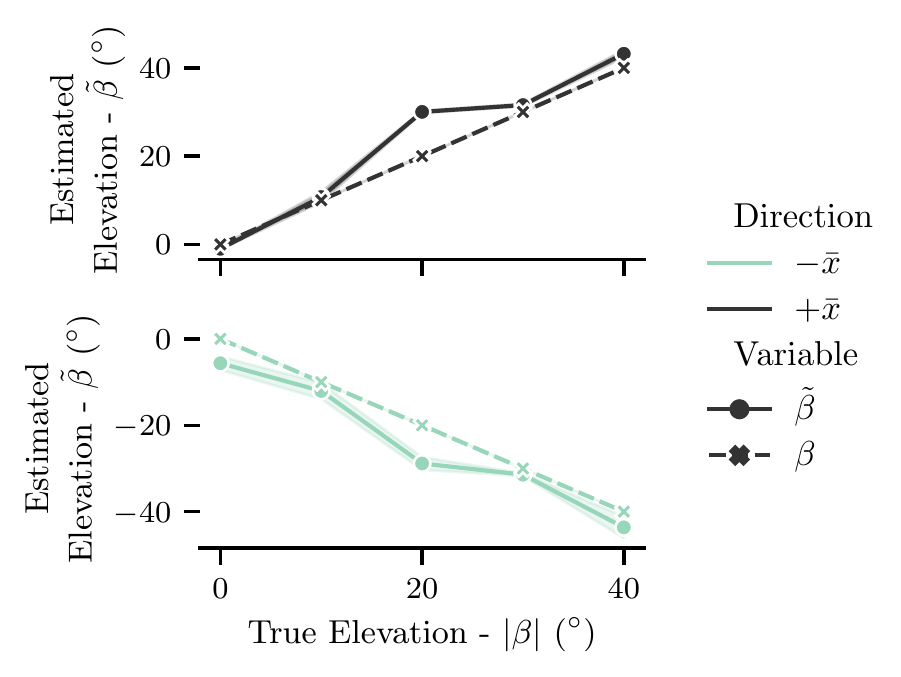}
  \caption{Estimated vs. true elevation angle $\beta$ in the tangential and radial velocity experiment. The solid lines represent the estimated values while the dashed lines represent the true values. The shaded region around the lines in the figures indicates the standard deviation of the 50 samples in each direction.}
  \label{elevation-angle}
\end{figure}

\begin{figure}[tb]
\includegraphics{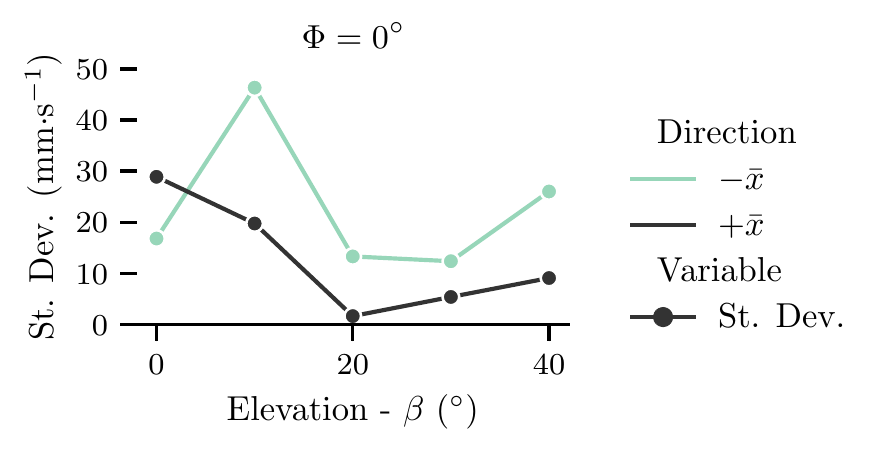}
\caption{\hlb{Measured target tangential velocity standard deviation on the $\Phi=\ang{0}$ baseline, at each elevation angle $\beta$.}}
\label{elevation-angular-crlb}
\end{figure}

\begin{figure}[tb]
  \includegraphics{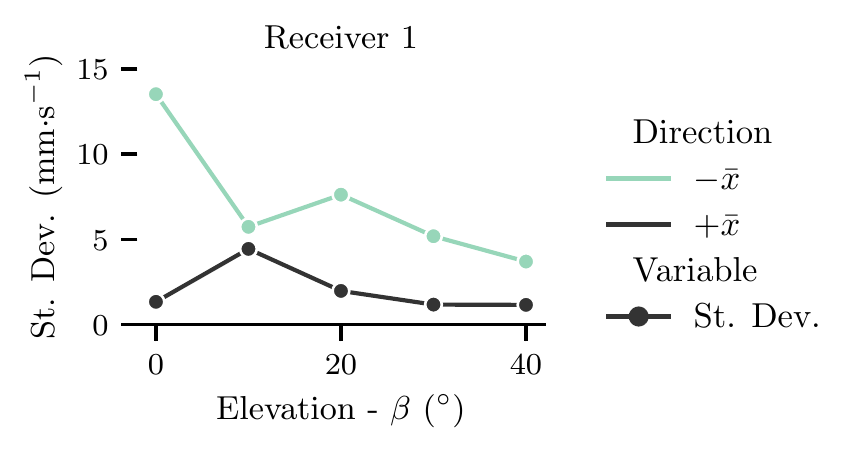}
  \caption{\hlb{Measured target radial velocity standard deviation on receiver 1, at each elevation angle $\beta$.}}
  \label{elevation-radial-crlb}
\end{figure}

%
%
%
%
%
%

\begin{table}[tb]
	\caption{Tangential and Radial Velocity Experiment Estimate Errors}
	\label{tab:ExII_error}
  	\begin{tabularx}{\linewidth}{*{7}{Y}}
	
	\toprule
	\multicolumn{2}{c}{Quantity} & Symbol & RMSE & Max. Err. \\

	\midrule
	
	\multicolumn{2}{c}{Radial Velocity (\SI{}{\milli\meter\cdot\second^{-1}})} & $\tilde{v}_R$ & \ExIIradVelRmse & \ExIIradVelMaxDev \\
	\multicolumn{2}{c}{Tangential Velocity (\SI{}{\milli\meter\cdot\second^{-1}})} & $\tilde{v}_\theta$ & \ExIItanVelRmse & \ExIItanVelMaxDev \\
	\multicolumn{2}{c}{True Velocity (\SI{}{\milli\meter\cdot\second^{-1}})} & $\|\tilde{\mathbf{V}}\|$ & \ExIItrueVelRmse & \ExIItrueVelMaxDev \\
	\multicolumn{2}{c}{Angle ($^\circ$)} & $\tilde{\beta}$ & \ExIIangRmse & \ExIIangMaxDev \\

	\bottomrule
	\end{tabularx}
\end{table}


The results of this experiment are summarized in \mbox{Figs. \ref{elevation-velocity} and \ref{elevation-angle}}. 
Fig. \ref{elevation-velocity} shows the total estimated velocity for the given angle of $\beta$ while Fig. \ref{elevation-angle} shows the estimated vs. actual elevation angles.  A summary of the estimation error for each of the quantities measured using the simple estimator described previously \hlb{is shown} in Table \ref{tab:ExII_error}.
The true velocity estimates shown in Fig. \ref{elevation-velocity} were produced using the $L^2$ norm of the radial and tangential velocities $\|\tilde{\mathbf{V}}\|=\sqrt{v_{\alpha_x}^2+v_R^2}$ where $v_{\alpha_x}$ is found using \eqref{3d_ang_vel} with $\Phi=\ang{0}$ and $v_R$ is found by estimating the Doppler shift on receiver 1. The \ac{rmse} for the \hlb{target's} true velocity was found to be \SI{\ExIItrueVelRmse}{\milli\meter\cdot\second^{-1}}.
Finally, the angle estimates in Fig. \ref{elevation-angle} were calculated using \eqref{beta}. Across all the measured elevation angles, an \ac{rmse} of \ang{\ExIIangRmse} was achieved.

Similarly to the first experiment, the three-dimensional speed estimate showed close agreement with the true speed of the linear guide, however the angle estimate had two notable errors: first at $|\beta|=\ang{0}$ in the $-\bar{x}$ direction, and the second at $|\beta|=\ang{20}$ in both directions.  The first is likely due to the positional estimation error of the target in calibrating for the offset from the center of the array at the time of measurement due to the direct-downconversion null, as previously described; due to a consistent underestimation of the positional offset when the guide was traveling in the $-\bar{x}$ direction, the Doppler component correction applied was not significant enough, causing a significant component of radial motion to be used when computing the elevation from \eqref{beta}.  The deviation at $|\beta|=\ang{20}$ is caused by a corresponding unexpected increase in the observed Doppler shift at that angle; it is possible that this may have been caused by a peak \ac{psd} in the observed Doppler spectrum manifesting at a point along the track at which the target had an elevation angle of greater than \ang{20}, causing the measured Doppler shift to be greater at that point.  This could be caused by several sources including multipathing, nonuniformities in the gain patterns of the 3-D printed horn antennas, or misalignment of the horns themselves due to minor deformation of the support material used to mount the antennas. 

\hlb{Finally, to provide a performance assessment of the system during this experiment, the standard deviation of the angular velocity estimate on the $\Phi=\ang{0}$ baseline and the Doppler velocity estimate on receiver 1 are presented in Figs. \ref{elevation-angular-crlb} and \ref{elevation-radial-crlb} respectively. The mean standard deviations for the correlation and Doppler measurements were \SI{16.11}{\milli\meter\cdot\second^{-1}} and \SI{4.78}{\milli\meter\cdot\second^{-1}} respectively, while the mean \acp{crlb} for the correlation and Doppler estimates were \SI{0.36}{\milli\meter\cdot\second^{-1}} and \SI{1.73}{\micro\meter\cdot\second^{-1}} respectively. The \ac{crlb} for the angular velocity was computed using \eqref{angular-velocity-accuracy} and the \ac{crlb} for the Doppler velocity was computed using \eqref{radial-velocity-accuracy}. Both \acp{crlb} used the estimated \ac{snr} from the measured data. As discussed in Section \ref{2d-tangential-experiment}, the discrepancy in magnitude at $|\beta|=\ang{0}$ is partly caused by the non-ideal positional estimate of the target off broadside when the peak power is received at the array, which is not incorporated into the \ac{crlb}, thus there is good potential to further improve the measurement accuracy. As the target angle increases, the effects of the DC-null diminish, however, the effects of the small-angle approximation begin to manifest, as well as uncertainty in the actual angle of the target off broadside at which the peak power is received. The estimator used in this paper assumes the peak power is received at broadside, however as the elevation angle $\beta$ increases, the effects of propagation loss may be greater than that of the antenna pattern, causing the returned power to be greater when the target is at a position off broadside; this would imply that the apparent baseline of the receiver is smaller than it would be at broadside, reducing the accuracy of the receiver.}

\section{Conclusion}

Experimental evidence for the efficacy of correlation interferometry for target velocity measurement across three dimensions has been presented.  
Using a simple estimator, clear correlations between the modeled and estimated values for the target trajectory are shown. 
In the experiments, measurement \acp{rmse} of \SI{\ExItanVelRmse}{\milli\meter\cdot\second^{-1}} for tangential measurements, and \SI{\ExIItrueVelRmse}{\milli\meter\cdot\second^{-1}} for measurements with a significant radial and tangential velocity were achieved. Three-dimensional speed measurements achieved \acp{rmse} of \ang{\ExItanAngRmse} and \ang{\ExIIangRmse} for fully tangential, and tangential and radial motion respectively, demonstrating the effectiveness of a dual axis correlation interferometer for the measurement of three-dimensional velocity.

\bibliographystyle{IEEEtran}
\bibliography{tmtt_2020.bib}

\end{document}